# Diffusion properties of electrons in GaN crystals subjected to electric and magnetic fields

G.I. Syngayivska[1], V.V. Korotyeyev[1], V.A. Kochelap[1]
  [1] V. Lashkaryov Institute of Semiconductor Physics,
  NAS of Ukraine, 41, prospect Nauky, 03028 Kyiv, Ukraine,
  e-mail: singg@ukr.net, koroteev@ukr.net

**Abstract**. We studied the diffusion coefficient of hot electrons of GaN crystals in moderate electric (1...10 kV/cm) and magnetic (1…4 T) fields. Two configurations, parallel and crossed fields, are analysed. The study was carried out for compensated bulk-like GaN samples at different lattice temperatures (30…300 K) and impurity concentrations ($10^{16}…10^{17}$ cm$^{-3}$). We found that at low lattice temperatures and low impurity concentrations, electric-field dependencies of the transverse-to-current components of the diffusion tensor are non-monotonic for both configurations, while the diffusion processes are greatly controlled by the magnetic field. With an increase of the lattice temperature or the impurity concentration, the behaviour of the diffusion tensor becomes more monotonous and less affected by the magnetic field. We showed that such behaviour of the diffusion processes is due to the distinct kinetics of the hot electrons in polar semiconductors with strong electron-optical phonon coupling. We suggest that measurements of the diffusion coefficient of the electrons subjected to electric and magnetic fields facilitate the identification of features of different electron transport regimes and the development of more efficient devices and practical applications.

## *I. Introduction*

The results of intensive investigations of wide-bandgap semiconductor compounds, in the particular, group-III nitrides, such as GaN, InN, AlN and related quantum heterostructures, can find various applications in the modern high-power and high-frequency microelectronics and optoelectronics [1, 2]. The nitride compounds are discussed as perspective materials for new devices such as light-emitting diodes [3, 4], optical switches [5], biosensors [6] and THz-active devices. The latter include electrically pumping THz sources [7-9], detectors [10] and modulators [5, 11]. A lot of attention is also paid to the electron transport properties of the nitrides in magnetic fields to develop novel devices working as sensors and switches controlled by a magnetic field [12-14].

In contrast to conventional $A^{III}B^{V}$ materials, such as GaAs, InSb or InP, the wide-bandgap nitrides (for GaN the bandgap is 3.2 eV) are described by the large separation between the lower $\Gamma$-valley and upper valleys separation (~1.2…1.5 eV for GaN), the high optical phonon energy, $\hbar\omega_0$ (for GaN, $\hbar\omega_0 \approx 92$ meV), the strong electron-optical-phonon coupling (the Fröhlich constant is ~0.4 for GaN) and the high low-field mobility (at room to liquid nitrogen temperatures the mobility is ~1500…5000 cm$^2$/V·s for GaN, [15]). These material properties of the nitrides are favorable for the realization of a specific streaming-like electron transport regime, which is characterized by a quasi-periodic electron motion in the momentum space [16-20] due to the threshold character of the electron-optical phonon emission. The streaming transport regime is possible at low lattice temperatures, $T_0$, ($k_B T_0 < \hbar\omega_0$, where $k_B$ is the Boltzmann constant) and small electron concentrations, $N_e$. The latter means that electron-electron scattering does not control the electron kinetics.

In papers [19, 20] it was shown that conditions of the streaming transport regime can be realized in the compensated high-quality bulk GaN samples. Under streaming regime, both steady-state and high-frequency characteristics have specific behaviour. In the particular, the current-voltage characteristics show a saturation, while field dependences of the diffusion coefficient demonstrate strongly non-monotonic behaviour. In addition, a high-frequency conductivity of the electrons is essentially anisotropic. Spectra of the high-frequency conductivity along the steady-state field have the oscillating behavior with appearance of the frequency "windows" with negative values of the real part of the conductivity. This effect is known as the optical phonon transit-time resonance (OPTTR). The OPTTR is considered as a perspective mechanism for electrically-pumped amplifiers and generators in the THz frequency range [8, 9]. The streaming regime and OPTTR can be identified in the optical experiments by the methods of THz-Fourier- or time-domain spectroscopies. The peculiarities of the THz transmission/absorption spectra for the GaN sample under conditions of the OPTTR were analyzed in [21].

Additional useful information about the streaming-like transport regime can be obtained investigating the galvano-magnetic characteristics at different configurations of the electric, $E$, and magnetic, $H$, fields. For compensated GaN, in crossed configuration of $E$ and $H$, it was found the strong effect of the magnetic field on the streaming-like electron distribution function. In particular, it was shown that in the range of the moderate electric (3…10 kV/cm) and magnetic (1.5…5 T) fields, the electron transport occurs in the form of a vortex-like motion in the momentum space. At higher magnetic fields, the effect of a collapse of the dissipative current occurs due to the strong suppression of the electron-optical phonon emission [22]. The field dependencies (vs $E$ and $H$) of the dissipative current, the Hall current, and the Hall electric field were studied in details for the compensated GaN [23].

This paper continues investigations of the hot electron transport in compensated GaN with a focus on the diffusion processes of the electrons in the real space. These processes are actual for systems with nonuniform electron concentrations and are described by the Fick law:

$$J_i = -eD_{ij}\frac{\partial N_e}{\partial x_j}, \qquad (1)$$

where $J_i$ is $i$-projection of the diffusion flux density, $D_{ij}$ is the diffusion coefficient tensor. Below we will study the $E$- and $H$- field dependences of $D_{ij}$.

It is well-known that under equilibrium conditions ($E = 0$, $H = 0$) in cubic crystals, the diffusion processes are characterized by a scalar diffusion coefficient, $D$, which obeys to Einstein relationship

$$D = \mu k_B T_0 / e, \qquad (2)$$

where $\mu$ is the low-field electron mobility and $e$ is an elementary charge. This relationship is valid for non-degenerate electrons. For the case of the weak applied electric fields, when the electron distribution function remains quasi-isotropic, the diffusion coefficient can be obtained by solving the Boltzmann transport equation and can be expressed through the integrals of a steady-state distribution function in the momentum space [24, 25]. Under strongly non-equilibrium conditions two approaches for calculations of the diffusion coefficient are used. Both are based on the Monte Carlo simulation of the electron transport. The first approach is based on calculations of the variance of random distances of travels of an individual electron per unit time [26, 27]. The obtained diffusion coefficient describes the spreading of initially localized in space electron packet, i.e., it directly corresponds to the Fick law. The second approach uses the computation of the velocity autocorrelation function in the time domain [27, 28]. The both approaches give the same results in the case of the absence of the electron-electron correlation [29].

The experimental techniques of the diffusion coefficient measurements are based on the electro-gradient measurements of the thermo-electric force [30], time-of-flight measurements [31], noise spectroscopy [32], light-induced grating technique [33, 34], etc.

The mentioned above theoretical and experimental studies of diffusion properties of the non-equilibrium electron gas are dated by 1970-1980-th and were concentrated to conventional semiconductors such as Si, Ge, GaAs, etc. In the past decades, the theoretical studies of the diffusion properties of the hot electrons were oriented on the nitride compounds, GaN [35] and InN [36]. Particularly, in these papers, the electric field dependencies of the diffusion coefficient were obtained for the wide range of the electric field: $E = 0...100$ kV/cm. The non-monotonic behaviour of the diffusion coefficient was identified for both longitudinal and transverse diffusion with respect to the electric field direction.

The present interest to the study of the diffusion processes in a non-equilibrium electron gas is inspired, in the particular, by the development recent pure all-optical pump-probe techniques, known as the light-induced *transient* grating (LITG) technique [37]. This contactless technique allows to determine the diffusion coefficient as well the carrier recombination times by investigating temporal changes of the diffraction efficiency of the grating induced optically near the surface of the material.

The aim of this paper is the theoretical study of the diffusion processes of the hot electrons in compensated GaN subjected to the electric and magnetic fields of moderate magnitudes. The paper is organized as follows. In Section II, we briefly describe the model of the electron transport and the Monte Carlo method which used for the determination of the diffusion tensor. In section III, we discuss the steady-state electric characteristics for different samples of GaN at zero magnetic field. Then, the effect of the magnetic field on the diffusion coefficient are considered for the parallel and crossed configurations of electric and magnetic fields sections IV and V, respectively. In Section VI, we summarize the main results of these investigations.

## *II.    The model of the electron transport in electric and magnetic fields*

We consider the electron transport in the samples of GaN of cubic modification under the parallel and crossed configurations of electric, $E$, and magnetic, $H$, fields. At $E \parallel H$, we assume that $E$ and $H$ are directed along $z$-axis (Fig. 1 (a)). In the case $E \perp H$, it is supposed that $E$ and $H$ are applied along $z$- and $y$-axis respectively (Fig. 1 (b)). Here, the GaN-samples are assumed to be with short-circuited Hall contacts.

**Figure 1 should be placed here**

For calculations of electron transport characteristics we used the single-particle algorithm of the Monte Carlo procedure [20, 22, 23, 38]. Three main scattering mechanisms were taken into account: electron scatterings by ionized impurities, acoustic phonons and polar optical phonons. We considered the range of the electric fields for which the intervalley transitions to the upper valleys are absent and the electron dispersion law can be set the parabolic one. The electron-electron scattering is not included into the calculations because only small electron concentrations were considered.

The standard single-particle Monte Carlo algorithm is based on the stochastic simulation of the electron trajectories in the momentum and coordinate spaces including the scattering processes (for details, see [39, 40]). The instantaneous values of the electron velocity (momentum and/or energy) and coordinate are recorded at the specific moments of time and accumulated for the further statistical processing. Motion of the electrons in the electric and magnetic fields were treated as semiclassical. These fields were taken into account at the modelling of the electron trajectory between sequential collisions; however the fields were not included to the scattering probabilities.

In the case of the parallel configuration of $E$ and $H$, the equations of the electron motion in momentum space are following:

$$\begin{cases} \dot{p}_x = \omega_c p_y \\ \dot{p}_y = -\omega_c p_x \\ \dot{p}_z = eE \end{cases} \quad (3a)$$

For the crossed configuration, they read as:

$$\begin{cases} \dot{p}_x = -\omega_c p_z \\ \dot{p}_y = 0 \\ \dot{p}_z = eE + \omega_c p_x \end{cases} \quad (3b)$$

Here $p_x$, $p_y$, $p_z$ are the components of the electron momentum, $\omega_c$ is the cyclotron frequency, $\omega_c = eH/m^*c$, $m^*$ and $c$ are the effective electron mass and light velocity, respectively.

At the end of a free flight, the electron velocity is calculated by the following equations for parallel

$$\begin{cases} p_x(t) = p_x(t_0)\cos[\omega_c(t-t_0)] + p_y(t_0)\sin[\omega_c(t-t_0)] \\ p_y(t) = -p_x(t_0)\sin[\omega_c(t-t_0)] + p_y(t_0)\cos[\omega_c(t-t_0)] \\ p_z(t) = p_z(t_0) + eE(t-t_0) \end{cases} \quad (4a)$$

and crossed configuration

$$\begin{cases} p_x(t) = (p_x(t_0) + eE/\omega_c)\cos[\omega_c(t-t_0)] - p_z(t_0)\sin[\omega_c(t-t_0)] - eE/\omega_c \\ p_y(t) = p_y(t_0) \\ p_z(t) = (p_x(t_0) + eE/\omega_c)\sin[\omega_c(t-t_0)] + p_z(t_0)\cos[\omega_c(t-t_0)] \end{cases} \quad (4b)$$

Here $t_0$ and $t$ are respectively initial and final moments of a free flight.

The calculation of the electron coordinates are easily incorporated in the Monte Carlo scheme. The following equations are used in our Monte Carlo procedure for the simulation of the electron trajectory in the coordinate space:

$$\begin{cases} x(t) = x(t_0) + \dfrac{p_y(t_0)}{m^*\omega_c} + \dfrac{p_x(t_0)}{m^*\omega_c}\sin[\omega_c(t-t_0)] - \dfrac{p_y(t_0)}{m^*\omega_c}\cos[\omega_c(t-t_0)] \\ y(t) = y(t_0) - \dfrac{p_x(t_0)}{m^*\omega_c} + \dfrac{p_x(t_0)}{m^*\omega_c}\cos[\omega_c(t-t_0)] + \dfrac{p_y(t_0)}{m^*\omega_c}\sin[\omega_c(t-t_0)] \\ z(t) = z(t_0) + p_z(t_0)(t-t_0)/m^* + eE(t-t_0)^2/2m^* \end{cases} \quad (5a)$$

and

$$\begin{cases} x(t) = x(t_0) + \dfrac{p_x(t_0) + eE/\omega_c}{m^*\omega_c}\sin[\omega_c(t-t_0)] - \dfrac{p_z(t_0)}{m^*\omega_c} + \\ \qquad + \dfrac{p_z(t_0)}{m^*\omega_c}\cos[\omega_c(t-t_0)] - \dfrac{eE}{m^*\omega_c}(t-t_0) \\ y(t) = y(t_0) + p_y(t_0)(t-t_0)/m^* \\ z(t) = z(t_0) - \dfrac{p_x(t_0) + eE/\omega_c}{m^*\omega_c}\cos[\omega_c(t-t_0)] + \dfrac{p_x(t_0)}{m^*\omega_c} + \\ \qquad + \dfrac{p_z(t_0)}{m^*\omega_c}\sin[\omega_c(t-t_0)] + \dfrac{eE}{m^*\omega_c^2} \end{cases} \quad (5b)$$

Equations (5a) and (5b) were used for the determination of the electron coordinates in the case of the parallel and crossed configuration respectively.

For the calculation of components of the diffusion tensor, $D_{ij}$, we used the following equation [27, 39, 40]:

$$D_{ij} = \frac{1}{2}\frac{d}{dt}\left\langle (r_i(t)-\langle r_i(t)\rangle)(r_j(t)-\langle r_j(t)\rangle)\right\rangle \quad (6)$$

where $r_{i,j=1,2,3} = \{x, y, z\}$ are calculated according to equations (5a) or (5b). Angle brackets denote the time average.

In the zero magnetic field and for the parallel configuration of the fields, $E \parallel H$, the tensor of the diffusion coefficients has three non-zero diagonal components: $D_{xx}$, $D_{yy}$, $D_{zz}$ and $D_{xx} = D_{yy}$. (see E-H orientation on Fig. 1). At $E \perp H$, the tensor $D_{ij}$ have five non-zero components $D_{xx}$, $D_{yy}$, $D_{zz}$, $D_{xz}$ and $D_{zx} = D_{xz}$. Below we will analyze the components of the tensor $D_{ij}$ corresponding to the directions transverse to the electric field.

## *III. Electrical characteristics at H = 0*

In this section we discuss the electron transport characteristics including the drift velocity, $V_d$, the average energy, $\langle\varepsilon\rangle$, as well the diagonal components, $D_{xx} = D_{yy}$, of the diffusion tensor, all as functions of the electric field at the absence of the magnetic field. The parameters $D_{xx}$ and $D_{yy}$ describe the electron diffusion transversal to the applied field. We present the electric characteristics for three particular examples, which differ by the lattice temperature, $T_0$, the concentration of ionized impurities, $N_i$, and the electron concentration, $N_e$. For the case I we assume the following parameters: $N_i = 10^{16}$ cm$^{-3}$ $N_e = 10^{15}$ cm$^{-3}$ and $T_0 = 30$ K, for the case II – $N_i = 10^{17}$ cm$^{-3}$ $N_e = 10^{16}$ cm$^{-3}$ and $T_0 = 30$ K, and for the case III – $N_i = 10^{16}$ cm$^{-3}$ $N_e = 10^{15}$ cm$^{-3}$ and $T_0 = 300$ K. The features of the streaming transport regime are expected to be well-pronounced for the case I.

Fig. 2 demonstrates the total scattering probability, $W_{tot}(\varepsilon)$, as a function of the electron energy, $\varepsilon$, calculated for three discussed cases. The curves marked by I, II and III correspond to cases I, II and III, respectively. The larger difference between values of $W_{tot}$ in the passive ($\varepsilon < \hbar\omega_0$) and the active ($\varepsilon > \hbar\omega_0$) energy regions is preferable for a realization of the well-developed streaming transport regime (the case I). The large values of the total scattering probability in the active region correspond to the intensive optical phonon emission. These values are of two order larger than in the passive region, where the less intensive acoustic phonon and ionized impurity scatterings occur at $T_0=30$ K. The increasing of the impurity concentration (the case II) at a given temperature leads to the increasing of $W_{tot}(\varepsilon)$ in the passive region due to the increasing of the electron-impurity scattering. The increasing of $W_{tot}(\varepsilon)$ in the passive region with increasing of the temperature (the case III) at a given impurity concentration is associated with the activation of absorption of the optical phonons.

**Figure 2 should be placed here**

The field dependencies of the drift velocity, $V_d(E)$, the average electron energy, $\langle\varepsilon\rangle$, the average energy correspondent to the transversal electron motion, $\langle\varepsilon_\perp\rangle = (p_x^2 + p_y^2)/2m^*$, and the transverse component of the diffusion coefficient, $D_{xx}(E)$, are shown in Fig. 3. As expected, the characteristic features of the streaming regime are observed only for sample I. In the range of $E = 3…10$ kV/cm, the drift velocity and the average energy saturate (see Fig. 3(a), (b)), reach one half of a characteristic velocity $V_0 = \sqrt{2\hbar\omega_0/m^*}$ and the average energy approaches to $\hbar\omega_0/3$, respectively. As seen, for sample II with larger impurity concentration, a well-developed streaming regime is not formed. At the room temperature, in the range of $E = 3...10$ kV/cm, electron gas remains almost quasi-equilibrium, so $V_d(E)$ shows linear behaviour and $\langle\varepsilon\rangle(E)$ is close to its equilibrium value of $3/2\cdot k_B T_0$.

The emergence of the streaming regime can be clearly identified by the strong non-monotonic field dependence of the transverse component of the average electron energy, $<\varepsilon_\perp>$ (see Fig. 3 (c)). Such dependence is observed for the case I. The increasing of $<\varepsilon_\perp>$ at the $E$ up to ~1 kV/cm is associated with the initial heating of the electron gas. With further increasing of the field, $<\varepsilon_\perp>$ decreases due to the formation of a streaming-like distribution function elongated along field direction [20]. This decreasing tends to saturation in the fields of 3–10 kV. For the cases II and III, the streaming is not formed and $<\varepsilon_\perp>$ has a slightly nonmonotonic character.

**Figure 3 should be placed here**

The field dependences of the transverse-to-$E$ component of the diffusion coefficient, $D_{xx}(E)$, (shown on Fig. 3 (d)) are qualitatively similar to the dependences of $<\varepsilon_\perp>$. The specific strong nonmonotonic dependence of the $D_{xx}(E)$ is inherent for the streaming regime, that is realized for the sample I. At $E < 500$ V, the magnitude of $D_{xx}$ rapidly increases from the equilibrium value of 13 cm$^2$/s to the maximum of about 250 cm$^2$/s. The isotropic spreading in the transverse direction of the electrons in the momentum space is the main reason of such growth of the diffusion coefficient. The maximum of $D_x$ corresponds to the electric field at which the rapid spreading of the distribution function is terminated. At this field, the essential part of high energy electrons loses their energy due to the emission of the optical phonons. With further increasing of $E$, the streaming-like distribution function begins to form. At this, $D_{xx}$ rapidly decreases and at $E > 5$ kV/cm approaches to values of 20–25 cm$^2$/s. For the samples II and III, $D_{xx}(E)$ slowly decreasing from 50 cm$^2$/s to 30 cm$^2$/s with increasing of the $E$ from 1 kV/cm to 10 kV/cm.

The effect of the magnetic field on the behaviour of the diffusion coefficient is analyzed in the next sections.

## IV. Diffusion coefficient at the E || H configuration

It should be noted, that for electrons with parabolic dispersion law the application of the magnetic field along electric one has no affect on transport characteristics $V_d(E)$ and $<\varepsilon>(E)$ [38]. It follows from the equations (2a), (3a) from which it is seen that the electron motion in directions along and transverse to fields is uncoupled. However, the electron diffusion process in coordinate space shows the strong dependence on the magnetic field, $H$.

Fig. 4 demonstrates the electric field dependencies of the transverse-to current component of the diffusion tensor, $D_{xx}(E)$, calculated at three values of $H$. Panels (a), (b) and (c) correspond to the samples I, II and III respectively. For comparison, the component of $D_{xx}(E)$ at $H = 0$ is shown by the dashed curve. Curves 1, 2 and 3 correspond to three magnitudes of the magnetic field of 1.1 T, 2.3 T and 3.4 T, respectively. The behaviour of the dependence of $D_{xx}(E)$ in magnetic field has the following general peculiarities for all samples. (i) The dependences of $D_{xx}(E)$ are nonmonotonic with maximum which shifted to the higher electric fields with increasing of the $H$. (ii) The magnetic field suppresses the diffusion in transversal directions with respect to $E$ and $H$. (iii) The effect of the magnetic field decreases at higher electric fields.

**Figure 4 should be placed here**

The magnetic field effect on the electron diffusion is more essential for the sample I, for which the streaming regime is realized. Even at weak magnetic fields, it is observed a strong suppression of the maximum of diffusion coefficient, which decreases from value ~250 cm$^2$/s at $H = 0$ to $\approx 60$ cm$^2$/s at $H = 1.1$ T. With further increasing of the magnetic field, the maximum of

$D_{xx}$ progressively decreases and has the values of ≈ 30 cm$^2$/s at $H$ = 2.3 T and ≈ 15 cm$^2$/s at $H$ = 3.4 T. The position of the maximum of the diffusion coefficient correspond to the electric fields of 0.5, 1.5, 3 and 5.5 kV/cm for $H$ = 0, 1.1, 2.3 and 3.4 T, respectively. For the samples II and III, for which the streaming regime does not occurs, the magnetic field more weakly modifies the diffusion coefficient. For example, the maximum of $D_{xx}$ decreases only twice from ~60 cm$^2$/s at $H$ = 0 T to ~30 cm$^2$/s at $H$ = 3.4 T.

Fig. 5 allows us to compare the diffusion coefficient for the three samples at given magnetic fields. For the weak magnetic field ($H$ = 1.1 T), there is the range of the electric field, $E$ = 0...3 kV/cm, where the diffusion coefficient for sample I is larger than that for the samples II and III (see panel (a) on Fig. 5). At larger $H$, the $D_{xx}$ for sample I is smaller than $D_{xx}$ for samples II and III in the studied electric fields (see panels (b) and (c) on Fig. 5).

**Figure 5 should be placed here**

The results of our modelling of the diffusion processes in compensated GaN show that the largest variations of $D_{xx}(E)$ with the magnetic field occur in the samples, for which the streaming electron transport is formed. Both the electro-gradient and optical measurements (using LITG techniques) of the diffusion coefficient can provide additional experimental tools for identification of the streaming transport regime and hot electron parameters for GaN crystals.

## *V. Diffusion coefficient at the $E \perp H$ configuration*

In this section we consider the field dependencies of the components of the diffusion tensor corresponding to the transverse motion with respect to the electric field directions. In case $E \perp H$ assuming that $E$ and $H$ are oriented along $z$-axis and $y$-axis respectively (see Fig. 2 (b)), two components of the diffusion tensor, $D_{xx}$ and $D_{yy}$, will be analyzed. The $D_{xx}$ component describes the diffusion current perpendicular to $E$ and $H$. The $D_{yy}$-component characterizes the diffusion in the direction parallel to $H$ and perpendicular to $E$.

As discussed previously [22, 23, 41, 42], the magnetic field for this configuration strongly affects on the transport characteristics. It was established that the streaming transport regime can be destroyed by the magnetic field, forming a vortex-like distribution function in the momentum space. The result of the measurements of current-voltage characteristics depends on the form of the external circuits. Our transport model assumes the short-circuited Hall contacts. I

In case $E \perp H$ as well as $E \parallel H$, the behavior of $D_{xx}(E)$ depends on material parameters of the samples and the relation between the magnitudes of $E$ and $H$. The behavior of the component $D_{xx}$ has the same general features (i)-(iii), which are listed in previous section. As well seen from Fig. 6, (a), the impact of the magnetic field on the diffusion is largest for the sample I. The dependencies of $D_{xx}(E)$ show non-monotonic behavior with a maximum, which is shifted to the region of higher electric fields with increasing of $H$. However, at a given non-zero $H$, there is a region of the electric fields, where the diffusion is larger than at $H$ = 0. This peculiarity was absent in the case of $E \parallel H$. Moreover, the maximum of $D_{xx}(E)$ is more essentially shifted for the configuration $E \perp H$. For example, at $H$ = 1.1 T, the maximum of $D_{xx}$ is realized at $E$ = 3 kV/cm (for comparison, in parallel configuration at $E$ = 1.5 kV/cm). For the samples II and III (panels (b) and (c) in Fig 6, respectively) the effect of the magnetic field is weak, the dependencies of the $D_{xx}(E)$ are weakly non-monotonous with an extended region of the electric field $E$ = 1…6 kV/cm (at $H$ = 2.3…3.4 T), where $D_{xx}(E)$ is the almost constant.

**Figure 6 should be placed here**

In Fig. 7 the same dependencies of $D_{xx}(E)$ are grouped accordingly to given magnetic fields. As seen, at $H = 1.1$ T the $D_{xx}$-component calculated for the sample I exceeds the values of $D_{xx}$ obtained for samples II and III at the range of $E = 0.1…5$ kV/cm. With increasing of $H$, this range decreases and shifts to the higher electric fields. For the samples II and III, the field dependences of $D_{xx}$ are very similar in the wide ranges of $E$ and $H$.

**Figure 7 should be placed here**

The well-pronounced maximum of the $D_{xx}$-component of the diffusion tensor is realized for the sample I. It should be noted that the maximum arises in such fields $E$ and $H$ at which the formation of the specific form of steady-state distribution function in the momentum space occurs. This distribution function describes the magneto-transport regime with co-existence of two separated electron groups (for details see [22, 42]).

In contrast to the parallel configuration of $E$ and $H$, in the crossed configuration, components $D_{xx}$ and $D_{yy}$ are not equal. The $D_{yy}$ component describes a diffusion current along the magnetic field. The electric field dependencies of $D_{yy}$ at several values of $H$ are shown in Fig. 8. As seen, the behavior of $D_{yy}(E)$ cardinally differs from $D_{xx}(E)$. In general, the increase of $H$ leads to the growth of the electron diffusion in the $y$-direction for a wide range of the electric fields. Particularly, for the sample I (see panel (a) in Fig. 8) the amplitude and the position of the maximum weakly modify with increasing of the $H$ in the range 1.1 T…3.4 T. However, at weak magnetic fields, 0…1.1 T, the amplitude of the maximum of $D_{yy}$ is increased twice. With increasing of the electric field, the diffusion along the direction of the magnetic field progressively decreases and at $E > 6$ kV/cm it tends to the case corresponding to $H = 0$. For the sample II, the impact of the magnetic field on the dependence of $D_{yy}(E)$ is not essential so as for sample I. The maximum of $D_{yy}$ is increased by ~30% with increasing of the magnetic field from 0 T to 3.4 T. For the sample III, the magnetic field does not affect on the dependence of $D_{yy}(E)$.

**Figure 8 should be placed here**

We found that general behaviour of galvano-magnetic properties of the $D_{yy}$-component of the diffusion tensor correlates with behavior of the average energy of the electron motion in $y$-direction, $<\varepsilon_y>$. The results of our calculations of the electric field dependencies, $<\varepsilon_y>$, for the three samples are shown in Fig. 9 at three values of $H$.

**Figure 9 should be placed here**

## VI. *Summary*

We have studied the diffusion properties of the hot electrons in bulk-like GaN samples for parallel and crossed configurations of the applied electric and magnetic fields. It has been analyzed the field dependencies of the transverse components of the diffusion tensor which correspond to the electron motion in the directions perpendicular to the electric field. These results have been obtained by the Monte Carlo method of calculations. It was analyzed three types of GaN sample with a degree of the compensation of 90 %. It was assuming $T_0 = 30$ K, $N_i = 10^{16}$ cm$^{-3}$ for a sample I, $T_0 = 30$ K, $N_i = 10^{17}$ cm$^{-3}$ for a sample II and $T_0 = 300$ K, $N_i = 10^{16}$ cm$^{-3}$ for a sample III). Parameters of the sample I satisfy the requirements of the realization of the streaming transport regime.

We found that a strong impact of the magnetic fields on diffusion properties of the electron gas takes place for the sample I. In the parallel configuration of *E* and *H*, the electric field dependencies of the transverse-to-current components of the diffusion tensor have non-monotonous behavior with a maximum. The amplitude and the position of the maxima depend on the magnitudes of the magnetic field and are changed with the increasing of the magnetic field: the maximum is decreased and its position is shifted to the higher electric fields.

In the crossed configuration of *E* and *H*, the transversal-to-fields component of the diffusion tensor has similar behaviour. However, the positions of the maximum are more sensitive to the variation of *H*.

The suppression of the electron diffusion in transversal direction to *H* with the increase of the magnetic field is a general phenomenon, which is observed in both configurations. However, in the crossed configuration, the magnetic field enhances the electron diffusion along *H* direction. The electric field dependences of the longitudinal-to-*H* component of the diffusion tensor also have the maximum. Both position and amplitude of the maximum weakly depend on the magnitudes of *H*.

The parameters of the samples II and III prevent the formation of the streaming transport regime. This is the main reason of a weak affect of electric and magnetic fields on diffusion properties of the electrons in these samples. However, the main peculiarities observed in the field dependencies of the diffusion processes for sample I take place as well for the samples II and III.

We suggest that streaming transport regime and related magneto-transport effects can be investigated by measurements of the diffusion effects of the hot electrons in the electric and magnetic fields. Both the electro-gradient and optical measurements (using LITG techniques) of the diffusion coefficient can be used for this purpose. Also, the knowledge of high-field dependencies of the diffusion coefficient is important for the modeling of micron-scale electronic devices operated at strong electric and magnetic fields.

## Acknowledgments
This work is partially supported by the Ministry of Education and Science of Ukraine (Project M/24-2018) and German Federal Ministry of Education and Research (BMBF Project 01DK17028).

# References

1. J. Millan, P. Godignon, X. Perpina, A. Perez-Tomas and J. Rebollo, A survey of wide bandgap power semiconductor devices. *IEEE Transactions on Power Electronics*. 2014. **29,** No 5. P. 2155–2163.
2. S. Chowdhury, B. L. Swenson, M. H. Wong and U. K. Mishra, Current status and scope of gallium nitride-based vertical transistors for high-power electronics application. *Semicond. Sci. Technol*. 2013. **28,** No 7. P. 074014.
3. V. Avrutin, S. A. Hafiz, F. Zhang, Ü. Özgür, H. Morkoc and A. Matulionis, InGaN light-emitting diodes: Efficiency-limiting processes at high injection. *J. Vac. Sci. Technol. A*. 2013. **31,** No 5. P. 050809.
4. N. Lu and I. Ferguson, III-nitrides for energy production: photovoltaic and thermoelectric applications. *Semicond. Sci. Technol*. 2013. **28,** No 7. P. 074023.
5. M. Beeler, E. Trichas and E. Monroy, III-nitride semiconductors for intersubband optoelectronics: a review. *Semicond. Sci. Technol*. 2013. **28, No** 7. P. 074022.
6. V. A. Sydoruk, I. Zadorozhnyi, H. Hardtdegen, H. Luth, M. V. Petrychuk, A. V. Naumov, V. V. Korotyeyev, V. A. Kochelap, A. E. Belyaev and S. A. Vitusevich, Electronic edge-state and space-charge phenomena in long GaN nanowires and nanoribbons. *Nanotechnology*. 2017. **28**. P.135204.
7. K. Ahi, Review of GaN-based devices for terahertz operation. *Optical Engineering.*2017. **56,** No 9, P.090901.

8. V.V. Korotyeyev, V.A. Kochelap, K.W. Kim, and D.L.Woolard, Streaming distribution of two-dimensional electrons in III-N heterostructures for electrically pumped terahertz generation. *Appl. Phys. Lett*. 2003. **82,** No 16, P.2643–2645.
9. K.W. Kim, V.V. Korotyeyev, V.A. Kochelap, A.A. Klimov, and D.L. Woolard, Tunable terahertz-frequency resonances and negative dynamic conductivity of two-dimensional electrons in group-III nitrides. *J. Appl. Phys*. 2004. **96**, No 11, P. 6488–6491.
10. W. Knap, V. Kachorovskii, Y. Deng, S. Rumyantsev, J.-Q. Lu, R. Gaska, M. S. Shur, G. Simin, X. Hu, M. A. Khan, C. A. Saylor and L. C. Brunel. Nonresonant detection of terahertz radiation in field effect transistors. *J. Appl. Phys*. 2002. **91**, No 11, P. 9346–9353.
11. T. Laurent, R. Sharma, J. Torres, P. Nouvel, S. Blin, L. Varani, Y. Cordier, M. Chmielowska, S. Chenot, J.-P. Faurie, B. Beaumont, P. Shiktorov, E. Starikov, V. Gruzinskis, V. V. Korotyeyev and V. A. Kochelap. Voltage-controlled sub-terahertz radiation transmission through GaN quantum well structure *Appl. Phys. Lett*. 2011. **99.** P.082101.
12. G. Santoruvo, A. Allain, D. Ovchinnikov and E. Matioli, Magneto-ballistic transport in GaN nanowires. *Appl. Phys. Lett.* 2016. **109**. P.103102.
13. L. Bouguen, S. Contreras, A.B. Jouault, L. Konczewicz, J. Camassel, Y. Cordier, M. Azize, S. Chenot and N. Baron. Investigation of AlGaN/AlN/GaN heterostructures for magnetic sensor application from liquid helium temperature to 300 °C. *Appl. Phys. Lett*. 2008. **92.** P.043504.
14. A.M. Gilbertson, D. Benstock, M. Fearn, A. Kormanyos, S. Ladak, M.T. Emeny, C.J. Lambert, T. Ashley, S.A. Solin and L.F. Cohen. Sub-100-nm negative bend resistance ballistic sensors for high spatial resolution magnetic field detection. *Appl. Phys. Lett*. 2011. **98**, No 6. P. 062106.
15. D.C. Look and J.R. Sizelove, Predicted maximum mobility in bulk GaN // *Appl. Phys. Lett*. 2001. **79**, No 8. P.1133–1135.
16. E. Starikov, P. Shiktorov, V. Gruzinskis, L. Varani, C. Palermo, J-F Millithaler and L. Regiani, Frequency limits of terahertz radiation generated by optical-phonon transit-time resonance in quantum wells and heterolayers. *Phys. Rev. B*. 2007. **76, No** 4. P. 045333; Terahertz generation in nitrides due to transit-time resonance assisted by optical phonon emission. *J. Phys.: Condens. Matter*. 2008. **20**, No 38. P. 384209.
17. J.T. Lu and J.C. Cao, Monte Carlo study of terahertz generation from streaming distribution of two-dimensional electrons in a GaN quantum well. *Semicond. Sci. Technol.* 2005. **20**, No 8. P. 829–833.
18. E.A. Barry, K.W. Kim, and V.A. Kochelap, Hot electrons in group-III nitrides at moderate electric fields. *Appl. Phys. Lett.* 2002. **80**, No 13. P. 2317–2319.
19. O. Yilmazoglu, K. Mutamba, D. Pavlidis and T. Karaduman, Measured negative differential resistivity for GaN Gunn diodes on GaN substrate. *Electronics Lett.* 2007. **43**, No 8, P. 480–482.
20. G.I. Syngayivska and V.V. Korotyeyev, Electrical and high-frequency properties of compensated GaN under electron streaming conditions. *Ukr. J. Phys*.2013. **58**, No 1. P. 40–55.
21. V.V. Korotyeyev, Peculiarities of THz-electromagnetic wave transmission through the GaN films under conditions of cyclotron and optical phonon transit-time resonances *Semiconductor Physics, Quantum Electronics & Optoelectronics*. 2013. **16**, No 1, P. 18–26.
22. G.I. Syngayivska, V.V. Korotyeyev, V.A. Kochelap and L. Varani, Magneto transport in crossed electric and magnetic fields in compensated bulk GaN. *J. Appl. Phys.* 2016. **120**. P. 095704.

23. G.I. Syngayivska and V.V. Korotyeyev, Electron transport in crossed electric and magnetic fields under the condition of the electron streaming in GaN. *Semiconductor Physics, Quantum Electronics & Optoelectronics*.2015. **18**, No 1. P. 79–85.
24. V.V. Mitin, V.A. Kochelap, M.A. Stroscio Quantum Heterostructures: Microelectronics and Optoelectronics. – New York: Cambridge University Press, 1999. – 642 p. – ISBN-13: 978-0521636353.
25. B.R. Nag, Diffusion equation for hot electron *Phys. Rev. B.* 1975. **11**, No 8. P. 3031–3036.
26. M.S. Gupta, Random walk calculation of diffusion coefficient in two-valley semiconductors *J. Appl. Phys.* 1978. **49**, No 5. P. 2837–2844.
27. R. Fauquembergue, J. Zimmermann, A. Kaszynski, E. Constant and G. Microondes, Diffusion and the power spectral density and correlation function of velocity fluctuation for electrons in Si and GaAs by Monte Carlo methods. *J. Appl. Phys.* 1980. **51**, No 2. P. 1065–1071.
28. D.K. Ferry and J.R. Barker, Generalized diffusion, mobility, and the velocity autocorrelation function for high-field transport in semiconductors. *J. Appl. Phys.* 1981. **52**, No 2, P. 818–824.
29. *Hot electron diffusion* (ed. by J. Pozhela) [in Russian], Vilnuis, Mokslas, 1981.
30. J.K Pozhela and K.K. Repshas, Thermoelectric force of hot carriers. *Phys. Stat. Sol.* 1968. **27**, No 2. P. 757–762.
31. J.G. Ruch and G.S. Kino, Transport properties of GaAs. *Phys. Rev.* 1968. **174**, No 3. P. 921–931.
32. J.P. Nougier, J. Comallonga and M. Rolland, Pulsed technique for noise temperature measurement. *J. Phys. E: Sci. Instrum.* 1974. **7**. P. 287–290.
33. H.J. Eichler, P. Gunter, D.W. Pohl, Laser-Induced Dynamic Grattings. Berlin: Springer-Verlag, Berlin Heidelberg, 1986.
34. J. Linnros and V. Grivickas, Carrier-Diffusion Measurements in Silicon with a Fourier-Transient-Grating Metod. *Phys. Rev.* B. 1994. **50.** P.16943–16955.
35. .E. Starikov, P. Shiktorov, V. Gruzinskis, L. Reggiani, L. Varani, J.C. Vaissiere and C. Palermo, Monte Carlo calculations of hot-electron transport and diffusion noise in GaN and InN. *Semicond. Sci. Technol.* 2005. **20**, No 3, P.279–285.
36. S. Wang, H. Liu, B. Gao and H. Cai, Monte Carlo calculation of electron diffusion coefficient in wurtzite indium nitride. *Appl. Phys. Lett.* 2012. **100**, P.142105.
37. R. Aleksiejūnas, Ž. Podlipskas, S. Nargelas, A. Kadys, M. Kolenda, K. Nomeika, J. Mickevičius and G. Tamulaitis, Direct Auger recombination and density-dependent hole diffusion in InN. *Scientific Reports*. 2018. **8**. P. 4621.
38. G.I. Syngayivska, V.V. Korotyeyev and V.A. Kochelap, High-frequency response of GaN in moderate electric and magnetic fields: interplay between cyclotron and optical phonon transient time resonances. *Semicond. Sci. Technol.* 2013. **28**, No 3. P. 035007.
39. P.J. Price, Calculation of hot electron phenomena. *Solid-State Electronics*.1978. **21.** P. 9–16.
40. C. Jacoboni and L. Reggiani, The Monte Carlo method for the solution of charge transport in semiconductors with applications to covalent materials. *Rev. Mod. Phys*.1983. **55**, No 3. P. 645-705.
41. I. I. Vosilius and I. B. Levinson, Optical phonon production and galvanomagnetic effects for a large-anisotropy electron distribution. *JETP*. 1966. **23**, No 6, P. 1104–1107; Galvanomagnetic effects in strong electric fields during nonelastic electron scattering *JETP*. 1967. **25**, No 4. P. 672–679.
42. V.A. Kochelap, V.V. Korotyeyev, G.I. Syngayivska and L. Varani, High-field electron transport in GaN under crossed electric and magnetic fields. *Journal of Physics: Conference Series.* 2015. **647**. P. 012050.

# Figures:

**Figure 1**

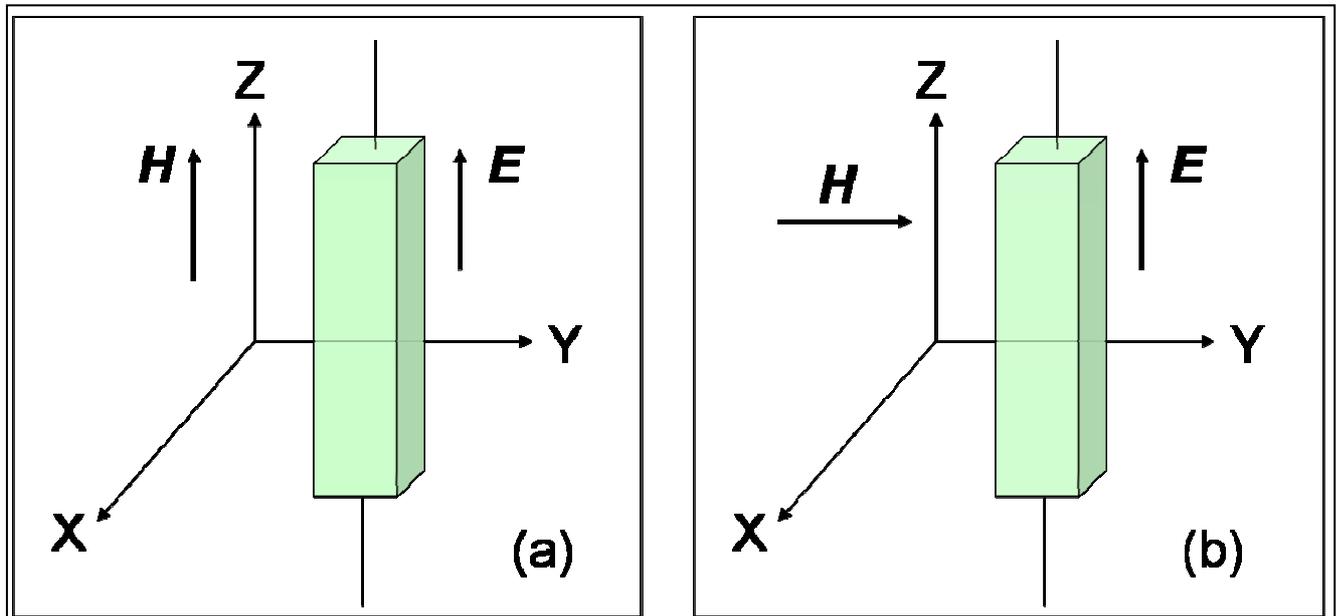

Fig. 1. The scheme of parallel (a) and crossed (b) configurations of *E* and *H*.

**Figure 2**

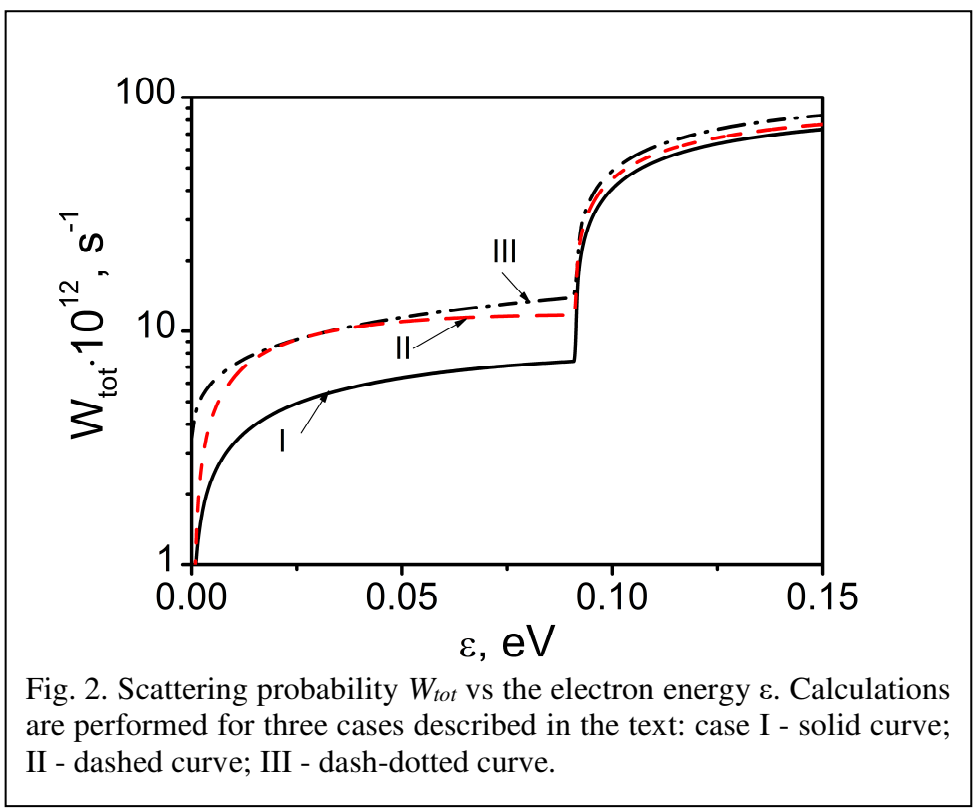

Fig. 2. Scattering probability $W_{tot}$ vs the electron energy $\varepsilon$. Calculations are performed for three cases described in the text: case I - solid curve; II - dashed curve; III - dash-dotted curve.

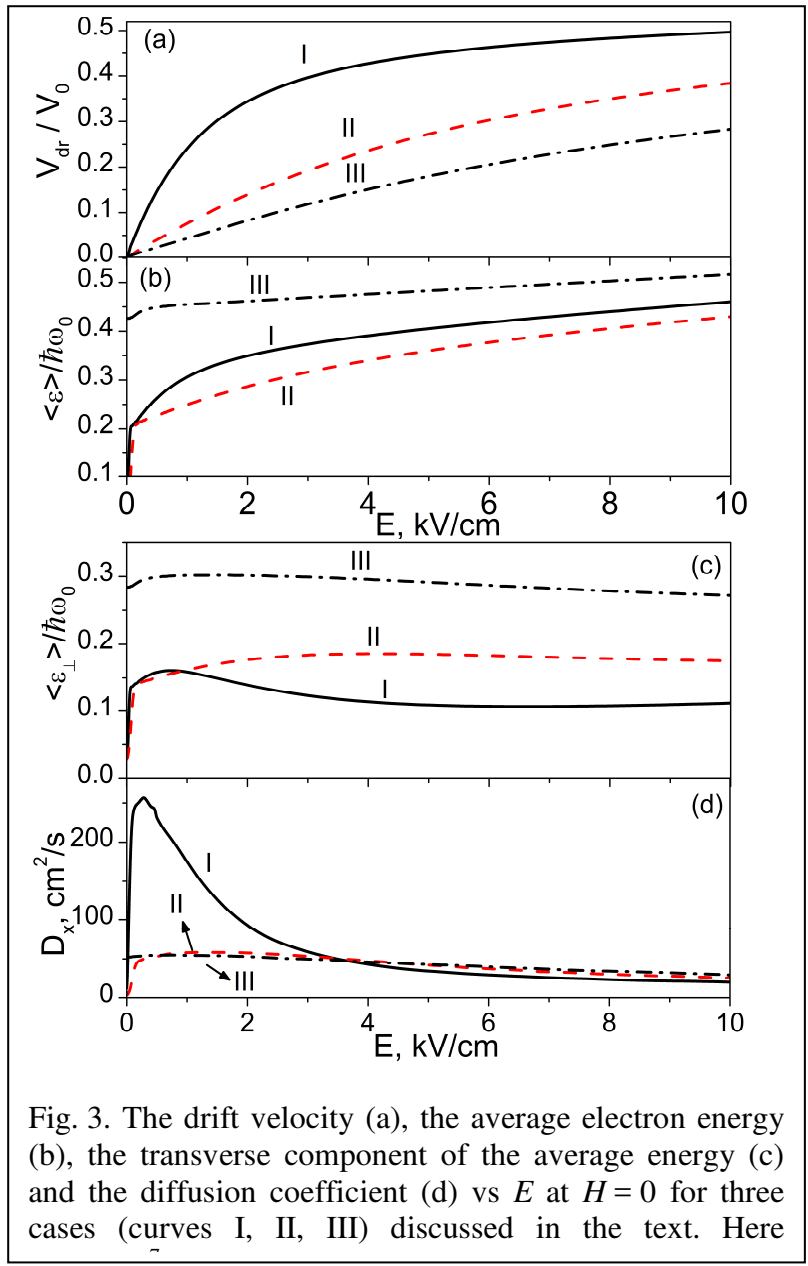

Fig. 3. The drift velocity (a), the average electron energy (b), the transverse component of the average energy (c) and the diffusion coefficient (d) vs *E* at *H* = 0 for three cases (curves I, II, III) discussed in the text. Here

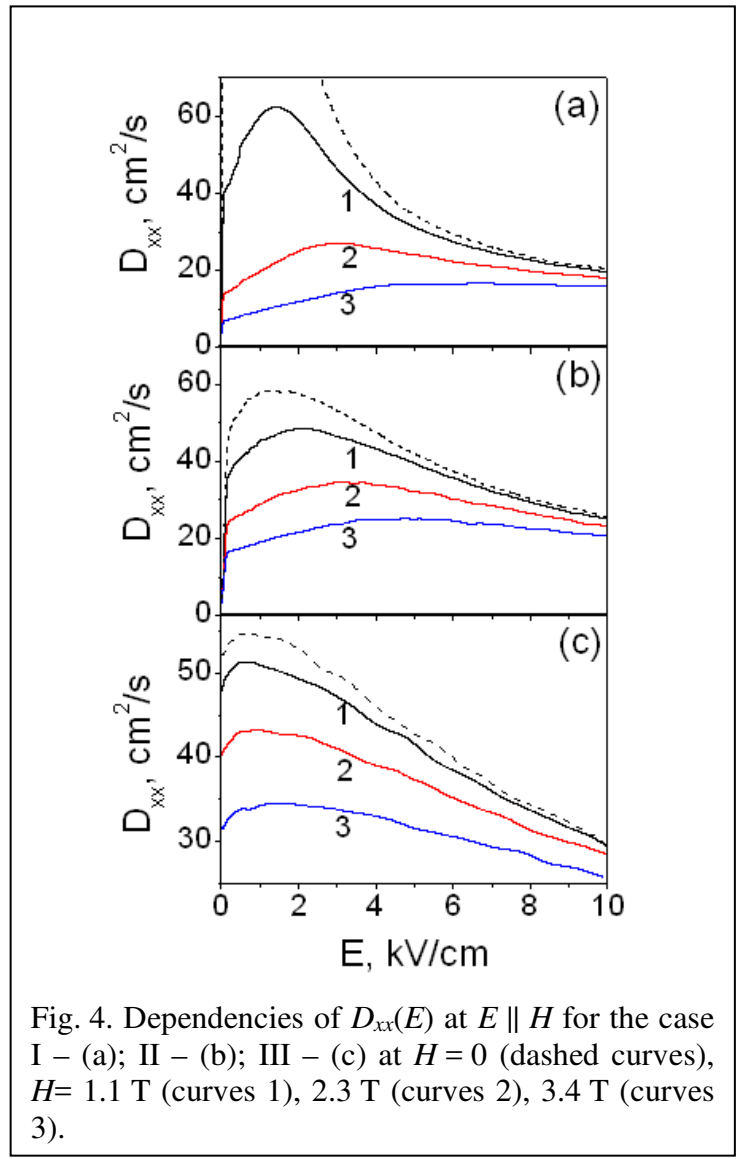

Fig. 4. Dependencies of $D_{xx}(E)$ at $E \parallel H$ for the case I – (a); II – (b); III – (c) at $H = 0$ (dashed curves), $H$= 1.1 T (curves 1), 2.3 T (curves 2), 3.4 T (curves 3).

**Figure 5**

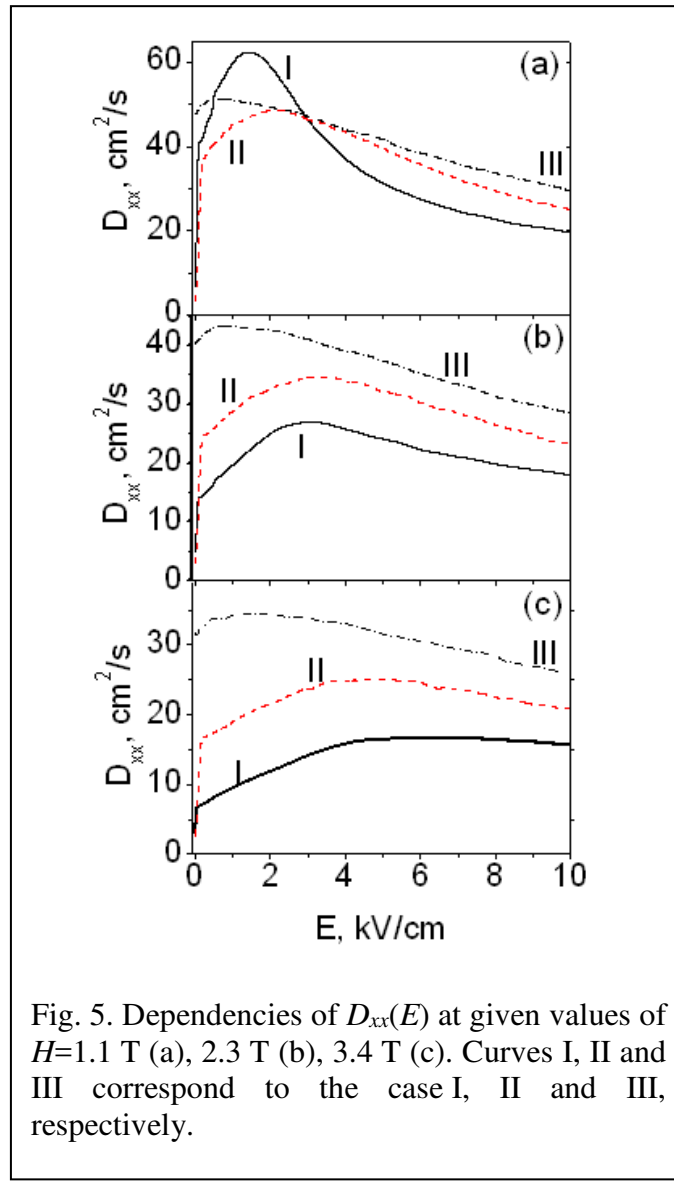

Fig. 5. Dependencies of $D_{xx}(E)$ at given values of $H$=1.1 T (a), 2.3 T (b), 3.4 T (c). Curves I, II and III correspond to the case I, II and III, respectively.

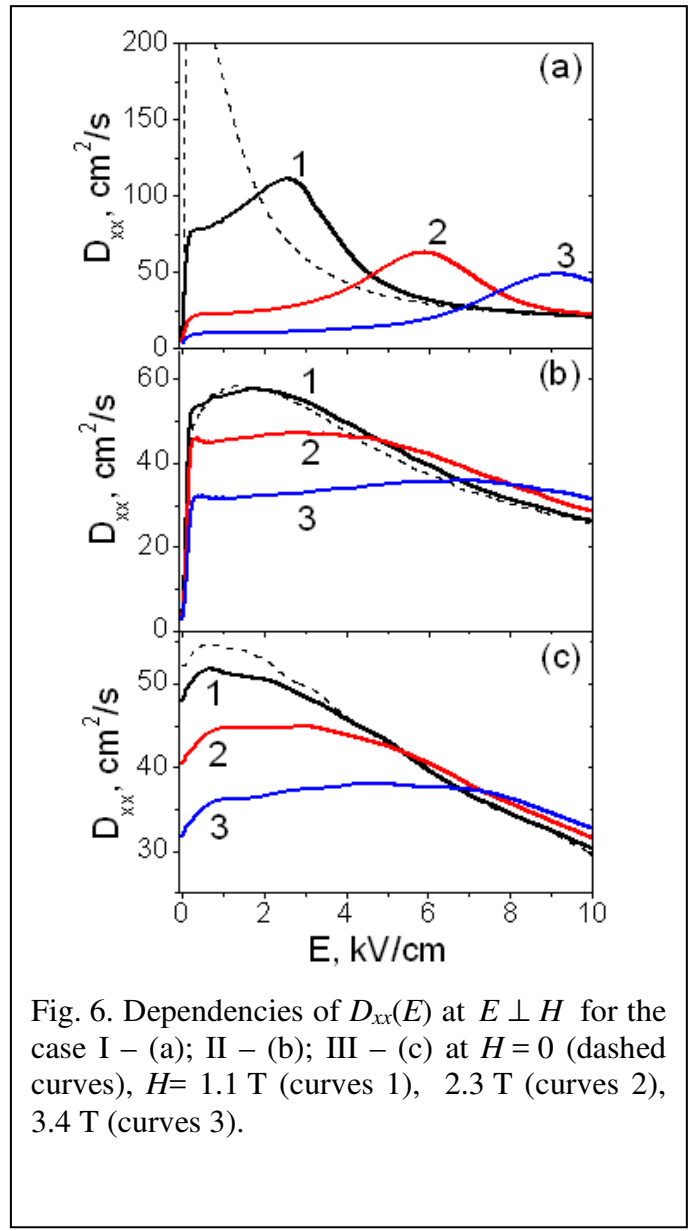

Fig. 6. Dependencies of $D_{xx}(E)$ at $E \perp H$ for the case I – (a); II – (b); III – (c) at $H = 0$ (dashed curves), $H= 1.1$ T (curves 1), 2.3 T (curves 2), 3.4 T (curves 3).

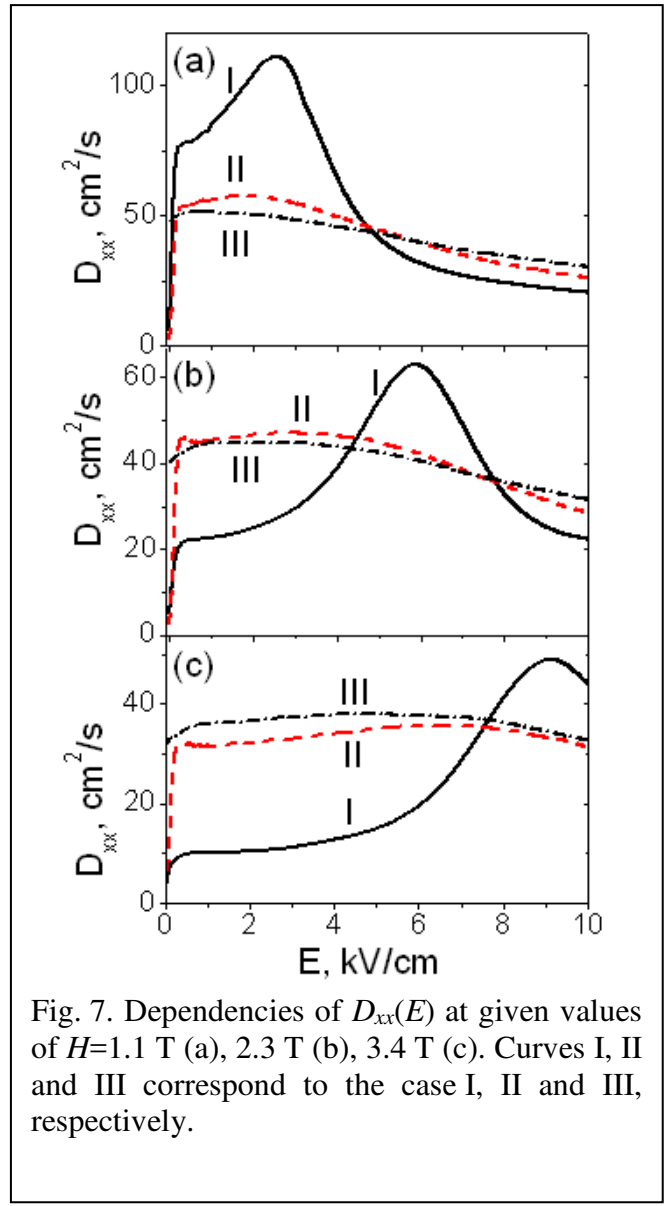

Fig. 7. Dependencies of $D_{xx}(E)$ at given values of $H$=1.1 T (a), 2.3 T (b), 3.4 T (c). Curves I, II and III correspond to the case I, II and III, respectively.

**Figure 8**

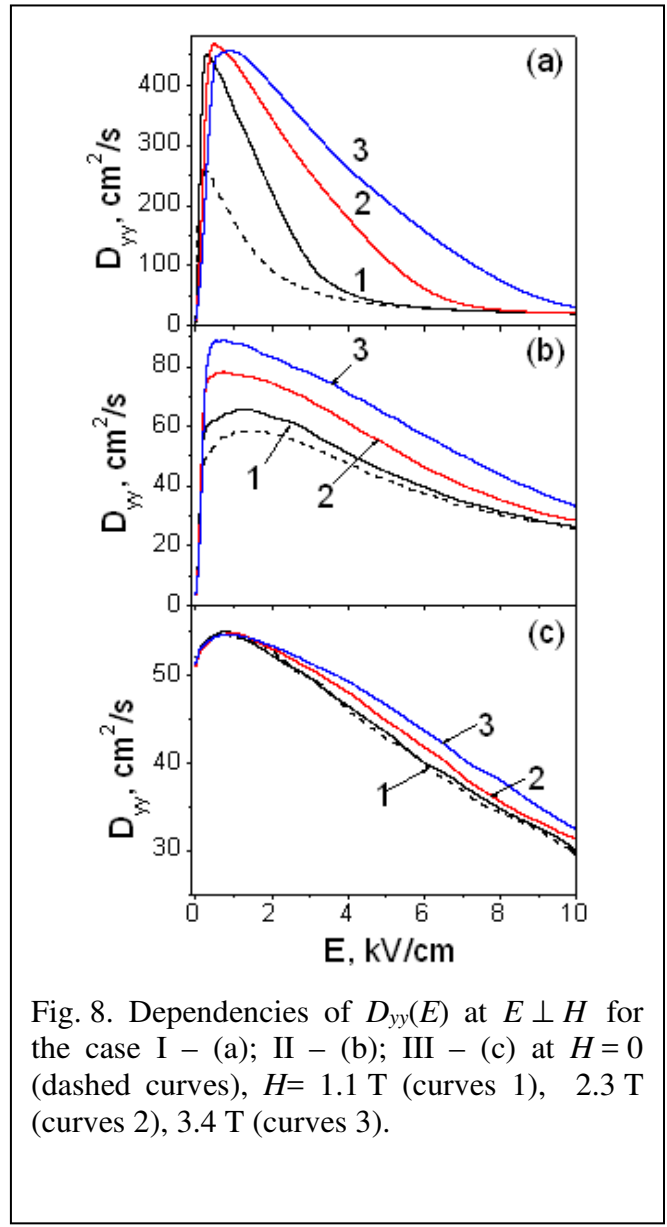

Fig. 8. Dependencies of $D_{yy}(E)$ at $E \perp H$ for the case I – (a); II – (b); III – (c) at $H = 0$ (dashed curves), $H= 1.1$ T (curves 1), 2.3 T (curves 2), 3.4 T (curves 3).

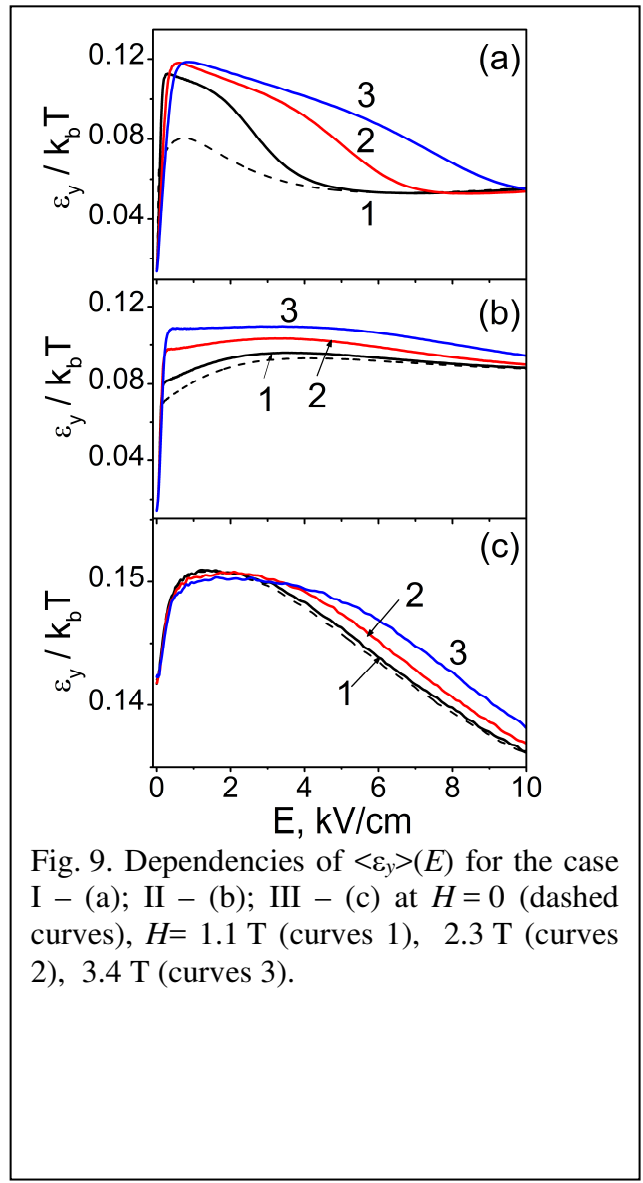

Fig. 9. Dependencies of $\langle\varepsilon_y\rangle(E)$ for the case I – (a); II – (b); III – (c) at $H = 0$ (dashed curves), $H = 1.1$ T (curves 1), 2.3 T (curves 2), 3.4 T (curves 3).